\documentclass[aps, prb,twocolumn,showpacs]{revtex4}
\usepackage{epsfig}
\usepackage{graphicx}
\usepackage{dcolumn}
\usepackage{bm}

\begin{document}

\title{Doping dependence of the chemical potential and surface electronic structure in
YBa$_2$Cu$_3$O$_{6+x}$ and La$_{2-x}$Sr$_x$CuO$_4$ using hard x-ray photoemission spectroscopy}

\author{Kalobaran Maiti,$^{1}$ J\"{o}rg Fink,$^{2,3}$ Sanne de Jong,$^4$ Mihaela Gorgoi,$^{2}$
Chengtian Lin,$^5$ Markus Raichle,$^5$ Vladimir Hinkov,$^5$ Michael Lambacher,$^6$, Andreas Erb,$^6$, Mark S. Golden,$^4$}

\affiliation{ $^1$ Department of Condensed Matter Physics and
Materials' Science, Tata Institute of Fundamental Research, Homi
Bhabha Road, Colaba,
Mumbai 400 005, India\\
$^2$ Helmholtz-Zentrum Berlin, Albert-Einstein-Strasse 15,12489 Berlin, Germany\\
$^3$ Leibniz-Institute for Solid State and Materials
Research Dresden, P.O.Box 270116, D-01171
Dresden, Germany\\
$^4$ Van der Waals-Zeeman Institute, University of Amsterdam, NL-1018XE Amsterdam, The Netherlands\\
$^5$ Max-Planck-Institute for Solid State Research, D-70569 Stuttgart, Germany\\
$^6$ Walther-Mei{\ss}ner-Institut, Bayerische Akademie der Wissenschaften, Walther-Mei{\ss}ner Strasse 8, 85748 Garching, Germany\\
}

\date{\today}

\begin{abstract}

The electronic structure of YBa$_2$Cu$_3$O$_{6+x}$ and
La$_{2-x}$Sr$_x$CuO$_4$ for various values of $x$ has been investigated using hard
$x$-ray photoemission spectroscopy. The experimental results
establish that the cleaving of YBa$_2$Cu$_3$O$_{6+x}$ compounds occurs predominantly
in the BaCuO$_3$ complex leading to charged surfaces at higher $x$ and to
uncharged surfaces at lower $x$ values.
The bulk component of the core level spectra exhibits a shift
in binding energy as a function of $x$, from which a shift of the chemical
potential as a function of hole concentration in the CuO$_2$ layers
could be derived. The doping dependence of
the chemical potential across the transition from a Mott-Hubbard insulator to a Fermi-liquid-like metal
is very different in these two series of compounds.
In agreement with previous studies in the literature the chemical potential shift in
La$_{2-x}$Sr$_x$CuO$_4$ is close to zero for small hole concentrations. In YBa$_2$Cu$_3$O$_{6+x}$, similar to
all other doped cuprates studied so far,  
a strong shift of the chemical potential at low hole doping is
detected. However, the results for the inverse
charge susceptibility at small $x$ shows a large variation between different doped cuprates. The
results are discussed 
in view of various theoretical models. None of these models turns out to be satisfactory.

\end{abstract}

\pacs{71.10.Hf, 71.30.+h, 73.20.-r, 74.25.Jb, 74.72.-h, 79.60.-i}

\maketitle

\section{Introduction}

High-$T_c$ superconductivity in cuprates\cite{Bednorz1986} arises from doping
of Mott-Hubbard insulators.\cite{Imada1998} The evolution of the electronic structure
as a function of doping concentration and the connected metal insulator transition
is still not well understood and may be the central
problem of high-$T_c$ superconductivity. Moreover the understanding of the metal-insulator
transition of doped Mott-Hubbard insulators as a material class is one of the outstanding challenges
occupying contemporary solid state science. Since in these correlated systems, and in particular in the doped
cuprates, there exists a complex
interplay between lattice, charge, and spin degrees of freedom, various phases
between the antiferromagnetic Mott-Hubbard insulator (or more precisely the charge-transfer insulator)
and the paramagnetic Fermi-liquid-like state have been the subject of discussion. The prominent state of
matter in this context is
the superconducting phase. In the underdoped region the pseudogap phase is also detected, which has been 
interpreted in terms
of preformed pairs,~\cite{Yang2008} a circulating current phase,~\cite{Chakravarty2001,Varma1999,
Simon2002,Fauque2006} or  exotic fractionalized states with topological order.~\cite{Sachdev2003,Lee2006,
Lee2008} In the underdoped region in many cuprates fluctuating or static 'stripe' phases 
play an important role. These phases
are essentially incommensurate unidirectional spin and charge density waves.\cite{Emery1999} 
In addition, stripe-like 'nematic' phases have been also predicted~\cite{Kivelson1998}
and finally observed in cuprates.~\cite{Hinkov2007,Hinkov2008} 

An interesting question which is related to the changes of the electronic structure of a doped 
Mott-Hubbard system near the metal-insulator transition is the evolution of the 
chemical potential $\mu$ as a function of 
electron density n. The rate of the chemical potential shift $\delta\mu/\delta n$
is equal to the inverse charge susceptibility $\chi_c^{-1}$ and is connected
to the static compressibility $\kappa$ of an electron liquid by the relation $\kappa=(1/n^2)(\delta n/\delta\mu)$.
Various scenarios have been discussed in relation to the question of the doping dependence of the
chemical potential shift in transition metal compounds. 
In a simple doped semiconductor model
the chemical potential should jump from the first affinity states  to the first ionization states 
when going from n-type 
doped to p-type doped systems.\cite{vanVeenendaal1993,Steeneken2003} In the cuprates, 
the former states correspond to the bottom of the upper Hubbard band while the latter 
are formed by the Zhang-Rice singlet 
states~\cite{Zhang1988} (top of the valence band). The energy difference between these states corresponds 
to the charge transfer gap which is
about 2 eV. Thus a jump of about 2 eV in $\mu$ would be expected when going from n to p-type doped 
systems or of about 1 eV when going from the
undoped system to the heavily p-type doped compound. An alternative scenario is that 
no shift in $\mu$ should occur as a function
of doping concentration due to a pinning of the chemical 
potential at impurity induced states which are in the gap.~\cite{Allen1990} 
A further model for a pinning of the
chemical potential is based on a 'microscopic' phase separation into a metallic and an insulating
phase, e.g. stripe or nematic phases.\cite{Emery1999,Vojta2009}

Various concrete predictions for the chemical potential shift as a function 
of the electron density n have been presented 
in the literature (see e.g. Ref.~\onlinecite{Georges1996}). For an isotropic Fermi liquid the relation for the rate 
of the chemical potential shift
\begin{equation}
\frac{\delta\mu}{\delta n}=(\frac{m_b}{m*})\frac{1+F_s^0}{N_b(\mu)}
\end{equation}
has been given,\cite{Furukawa1993} where m*, $m_b$, and $N_b(\mu$), are the effective mass, the bare band mass and the bare density 
of states (DOS) at the chemical potential $\mu$,
respectively. $F_s^0$ is a Landau parameter which represents the isotropic spin-symmetric part of the 
electron-electron interaction. In an uncorrelated system, $\delta\mu/\delta n$ would be just given
by the inverse bare density of state $1/N_b(\mu)$. 
In correlated systems the effective mass would strongly increase approaching the half-filled case. Thus a very small 
rate of chemical potential shift or a very large charge susceptibility $(\delta\mu/\delta n)^{-1}$ 
would be expected in this approximation.
A suppression of the chemical potential shift has been also predicted by numerical 
studies in the two dimensional Hubbard 
model\cite{Furukawa1993, Dagotto1991} for $n_b$ close to 1, where $n_b$ is the number of
electrons in the band. Here, from Monte Carlo studies a 
relation $\Delta\mu\propto-(n_b-1)^2$ 
has been arrived at. In addition, a slave boson mean-field analysis of the three-band model
yielded a constant chemical potential up to a hole 
doping $p$ = 1/8 due to a stripe-like phase separation.~\cite{Lorenzana2002} Finally, based on calculations in the $t-t'-t''-J$ model 
(which includes the long range hopping integrals $t'$ and $t''$), it has been predicted\cite{Tohyama2003} that 
$\delta \mu/\delta n$ should be small for $t'$=0 and should 
increase with increasing $t'$. 

Experimentally, the chemical potential shift can be deduced from core-level shifts in x-ray 
induced photoemission spectra (XPS). 
The reason for this is that the binding energy of the core level states is measured relative to the
chemical potential $\mu$. More explicitly, the shift in the binding energy of the core levels, $\Delta\epsilon$, depends
primarily on four terms and can be expressed as,
\begin{equation}
\Delta\epsilon = -\Delta\mu + \Delta{V_M} + K\Delta{Q} + \Delta{E_R}
\end{equation} 
where $\Delta\mu$, $\Delta{V_M}$, $\Delta{Q}$, K, and $\Delta{E_R}$
represent the change in chemical potential, the change in the
Madelung potential, the change in the valency, a constant, and the change in
relaxation energy, respectively.\cite{Huefner1995} It has turned out that in the 
cuprates and other transition metal compounds, for  core levels from
atoms which do not change their valency upon doping, 
the terms other than $\Delta\mu$ are small
and therefore the chemical potential shifts can directly derived from the core level shifts themselves.
This core level based method was first applied to
the system Bi$_2$Sr$_2$Ca$_{1-x}$Y$_x$Cu$_2$O$_{8+\delta}$ and a large total shift of the chemical potential
of $\approx$ 0.8 eV between undoped and overdoped samples was deduced, together 
with an inverse charge susceptibility
$\frac{\delta\mu}{\delta p}|_{p\to0}\approx$7 eV/hole for low hole 
concentration $p$.~\cite{vanVeenendaal1993}
For hole concentrations, p$\geq$0.1 this value was seen to decrease 
to $\approx$ 1 eV/hole.~\cite{vanVeenendaal1993}
Similar work has been 
performed on various other cuprates such as La$_{2-x}$Sr$_x$CuO$_4$ 
(Refs.~\onlinecite{Rietveld1995,Fujimori1996,Ino1997,Fujimori1998,Fujimori2002,Steeneken2003}), 
Nd$_{2-x}$Ce$_x$CuO$_4$ (Refs.~\onlinecite{Harima2001,Steeneken2003}), 
Bi$_2$Sr$_2$Ca$_{1-x}$R$_x$Cu$_2$O$_{8+\delta}$ 
(R = Pr, Er) (Refs.~\onlinecite{Tjernberg1997,Harima2003}),
Ca$_{2-x}$Na$_x$CuO$_2$Cl$_2$ (Ref.~\onlinecite{Yagi2006}), and 
Bi$_2$Sr$_{2-x}$La$_x$CuO$_{6+\delta}$ (Ref.~\onlinecite{Hashimoto2008}).
Since the electronic structure close to the Fermi level in all these compounds is determined
by doped CuO$_2$ layers, a universal behavior is expected. However, this is not observed for the
p-type doped cuprates, since La$_{2-x}$Sr$_x$CuO$_4$ shows  
a negligible $\frac{\delta\mu}{\delta p}|_{p\to0}$, but all other systems show a large 
$\frac{\delta\mu}{\delta p}|_{p\to0}\approx$1.8 eV/hole.~\cite{Hashimoto2008} This difference has often been 
explained in terms of a 'microscopic' stripe-like phase 
separation\cite{Vojta2009} and by the fact that 
La$_{2-x}$Sr$_x$CuO$_4$ or similar compounds in which Sr is replaced
by Ba, or La is replaced by Nd or Eu, are much more susceptible to static stripe formation than the other 
cuprates.~\cite{Harima2001,Fujimori2002} Alternative explanations of this puzzle have also been discussed 
in the literature, e.g. the influence of the opening of a pseudo gap on the doping dependence of the 
chemical potential.\cite{Fujimori1998} Finally, based on the above mentioned calculations in 
the $t-t'-t''-J$ model it was argued that for La$_{2-x}$Sr$_x$CuO$_4$,  
$\frac{\delta\mu}{\delta p}|_{p\to0}$ should be small due to its 
small value of $t'$, while 
$t'$ for the other cuprate superconductors is larger thus yielding larger 
$\frac{\delta\mu}{\delta p}|_{p\to0}$ values.

Introducing methods for the determination of chemical potential shifts, one should also mention 
that not only core-levels
but also the Cu 3d$^8$ satellite line\cite{Steeneken2003} and the lower 
Hubbard band\cite{Hashimoto2008} have been used as an 
intrinsic reference system for the determination of the doping dependence of $\mu$.

In this article we apply the core level binding energy method to study the chemical 
potential shift as a function of doping concentration 
in a further hole-doped cuprate, namely YBa$_2$Cu$_3$O$_{6+x}$ and compare 
the results with our own data for the chemical potential shift
of La$_{2-x}$Sr$_x$CuO$_4$. YBa$_2$Cu$_3$O$_{6+x}$ is an interesting system since it has 
not only two dimensional CuO$_2$ planes 
which are doped but also one dimensional CuO$_{2+x}$ units (see below),  
which transform into metallic ribbons
upon increasing the O concentration
from 6 to 7. Furthermore, in this context one should mention that YBa$_2$Cu$_3$O$_{6+x}$ is the only compound  
in which nematic stripe-like
order has been detected in the bulk.~\cite{Hinkov2008} 

YBa$_2$Cu$_3$O$_7$ forms in orthorhombic structure as shown in Fig.
1. As evident in the figure, Y layers are sandwiched by two CuO$_2$
planes ($a,b$-plane) whereby the latter are believed to be responsible for
superconductivity. The oxygens along the $a$ and $b$ axes in the CuO$_2$
planes are denoted by O(2) and O(3), respectively and are very similar.
The copper atoms in the CuO$_2$ planes are denoted by Cu(2) in the
figure. The apical oxygens that --- together with the Ba atoms form BaO layers --- are denoted by O(4)
and the oxygens in the CuO chains along the $b$ directions are denoted
by O(1). The O(1), O(4) atoms and the Cu(1) atoms together form a one dimensional CuO$_3$
metallic ribbon.
\begin{figure}
 \vspace{-2ex}
\includegraphics [scale=0.45]{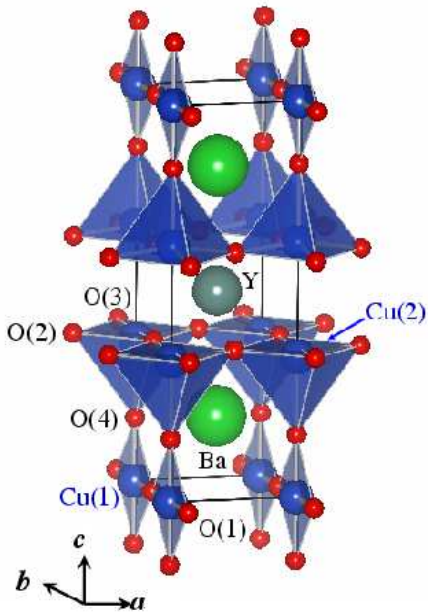}
\vspace{-8ex}
 \caption{(color online) crystal structure of YBa$_2$Cu$_3$O$_7$
(orthorhombic; space group $Pmmm$). O(1) sites are empty in the
crystal structure of YBa$_2$Cu$_3$O$_6$ leading to a tetragonal
structure.}
 \vspace{-2ex}
\end{figure}
The decrease in $x$ from $x$=1 to $x$=0 leads to the fragmentation of the
chains due to the absence of O(1) atoms and the orthorhombic structure
gradually changes near $x$=0.3 to a tetragonal structure.\cite{gallagher1987,Verweij1988}
Several studies have indicated ordering of oxygen atoms in the chain
and/or chain fragments.\cite{Jorgensen1990,Werder1988,Sonntag1991} 

The other end member, tetragonal YBa$_2$Cu$_3$O$_6$ is an
antiferromagnetic insulator. In this structure, the Cu(2) atoms are
formally divalent, which corresponds to a single hole in the Cu $3d$ shell. In order to
yield charge balance, the Cu(1) atoms need to be formally monovalent, 
i.e., their Cu $3d$ states are completely filled. Evidence of
such Cu$^{1+}$ atoms has been observed experimentally in various
studies, e.g. by x-ray absorption spectroscopy (XAS).\cite{Tranquada1988}  

Increasing $x$ from 0 to 1 , i.e., filling the O(1) sites, first the valency of the Cu(1) atoms
is changed from monovalent to divalent. This is completed near $x$=0.3. Further increase
of $x$ leads to the formation of formally trivalent Cu atoms in the chains and in the planes. Due
to the large on-site Coulomb interaction $U\approx$8 eV for two holes on a single Cu site
and the smaller charge transfer energy $\Delta\approx$3 eV, the holes are --- in reality ---
not formed at the Cu sites but reside on the adjacent O sites,~\cite{Nucker1988,Nucker1995} thus forming a 
Zhang-Rice singlet state together
with the already existing hole on the Cu sites.~\cite{Zhang1988} 
After an insulator to metal transition, 
the transition into the superconducting phase is subsequently observed. Interestingly, a plateau with superconducting
transition temperature $T_c$ of about 60 K for 0.6~$\leq x
\leq$~0.8 has been observed, which is significantly different from other
cuprates. Finally, a maximum $T_c$ of about 91 K is observed for $x$
close to 0.9.~\cite{Cava1987} 

As a reference system for YBa$_2$Cu$_3$O$_{6+x}$ we have studied La$_{2-x}$Sr$_x$CuO$_4$, which has the
K$_2$NiF$_4$ structure, where the CuO$_2$ planes are separated by two LaO
layers. La$_2$CuO$_4$ is an antiferromagnetic insulator. Divalent Sr
substitution at the trivalent La sites leads to hole doping,
thereby driving the system metallic. 
Higher doped compositions near $x$=0.15 exhibit superconductivity at about 40 K.

Photoemission is usually a highly surface sensitive method.\cite{Huefner1995} For photons
having an energy of 1.5 keV ( a commonly used x-ray energy in laboratory studies) 
the mean free path of the photoelectrons they create is of the order of
20~\AA. Increasing the photon energy to higher values reduces the surface sensitivity
and leads to results which are more closely related to bulk properties. In addition, the knowledge of
which lines are related to the surface and which are related to the bulk is of fundamental
importance for XPS studies on the chemical potential as a function of doping concentration. 
Therefore a prerequisite for our studies of the chemical potential
shift was a detailed investigation of the surface sensitivity of XPS spectra of YBa$_2$Cu$_3$O$_{6+x}$
using various photon energies and emission angles. 

Cleaving YBa$_2$Cu$_3$O$_{6+x}$ crystals perpendicular to the $c$ axis can be performed at 3 different planes
leading to 6 different possible surface terminations. At present it is not absolutely clear whether the crystals
are cleaving predominantly along one plane or whether other cleavage planes or even a mixture of the three
possible cleavage planes is occurring which may lead up to 6 different terminations on the same cleavage plane. 
The appearance of specific surface layers may also change as a function
of the doping concentration $x$.
Previously, XPS studies on the surface of YBa$_2$Cu$_3$O$_{6+x}$ have been published for 
single crystals,~\cite{Jorgensen1990,Fowler1990,Brundle1993} 
polycrystalline samples,~\cite{Steiner1987,Srivastava1993}
and thin films.~\cite{Frank1991,Teterin1993} In this context
we also mention scanning tunneling microscopy (STM)~\cite{Edwards1992,You1992,Derro2002,Maki2002} 
studies on single crystals. More recently, information on the surface of cleaved
YBa$_2$Cu$_3$O$_{6+x}$ single crystals have been also derived from 
Angle-Resolve Photoemission Spectroscopy (ARPES) studies.~\cite{Schroeder1993,Schroeder1993a,
Schabel1998,Schabel1998a,Kondo2007,Borisenko2006,Nakayama2007,Zabolotnyy2007,Zabolotnyy2007a,Hossain2008} 
Most of these investigations of the surface layers of cleaved single crystals
indicate that BaO and CuO layers are at the surface, i.e., the cleavage occurs predominately 
between the BaO layers and the one dimensional CuO chains. Some authors also
discussed a mixed CuO$_3$/BaO surface termination.~\cite{Schabel1998} On the other hand
in theoretical studies on the basis of density functional 
calculations (Refs.~\onlinecite{Calandra1992} and \onlinecite{Calandra1994}) it was pointed out that breakage
of the strong covalent Cu(1)-O(4) bond would be difficult.
STM sudies~\cite{You1992} also reported 
Y layers at the surface which would indicate a cleavage between the CuO$_2$ planes and the Y layers.
Finally, cleavage between the BaO and CuO$_2$ layers have been reported.~\cite{Schroeder1993}
Recent ARPES studies have also detected overdoped surfaces on cleavage of optimally doped bulk crystals.
Such surfaces with doping concentrations different from the bulk are expected 
when polar surfaces are produced after the cleavage process. 

In this article we present first a detailed XPS study of the electronic structure of cleaved YBa$_2$Cu$_3$O$_{6+x}$ 
and  La$_{2-x}$Sr$_x$CuO$_4$
single crystals. This is followed by the main focus of this contribution, 
the investigation of the chemical potential shifts
in these systems as a function of their CuO$_2$ plane doping concentration. 
  
\section{Experimental:}

High-quality, single-phase YBa$_2$Cu$_3$O$_{6+x}$ crystals were
synthesized by the solution growth method\cite{hinkov2004} or in nonreactive BaZrO$_3$ crucibles by a 
self flux method.\cite{Erb1996} The crystals were examined by various other
techniques such as inelastic neutron scattering,\cite{hinkov2004,Hinkov2008} muon 
spin rotation ($\mu$SR),~\cite{hinkov2004}
and ARPES.\cite{Borisenko2006,Zabolotnyy2007,Zabolotnyy2007a} 
In Table I we present the doping concentration $x$, the hole doping $p$ in the CuO$_2$ layers, 
the superconducting transition
temperatures $T_c$, and the widths of the superconducting transition temperature $\Delta T_c$. 
The non-trivial hole doping $p$ for CuO$_2$ layers was derived from a recent
$c$-axis lattice parameter study of YBa$_2$Cu$_3$O$_{6+x}$ crystals.~\cite{Liang2006} These values are in 
qualitative agreement with values
derived from XAS measurements on the O 1$s$ level on untwinned single crystals.~\cite{Nucker1995}
The hole doping concentration for the x=0.45 and 0.6 samples have been determined independently to be p=0.085
and 0.12, respectively.~\cite{Hinkov2008,Hinkov2007}  
La$_{2-x}$Sr$_x$CuO$_4$ crystals were grown with the traveling solvent zone method.~\cite{Lambacher2008}
The doping concentration of the crystals together with T$_c$ and $\Delta T_c$ are also listed in
Table I.
\begin{table}
\centering
\caption{Doping concentration $x$, hole doping p in the CuO$_2$ layers, $T_c$ and $\Delta T_c$ values of the
 YBa$_2$Cu$_3$O$_{6+x}$ (YBCO) and La$_{2-x}$Sr$_x$CuO$_4$ (LSCO) single crystals used in this work}
\label{tab:1}       
\begin{tabular}{lllrr}
\hline\hline\noalign{\smallskip}
&\hspace{1cm}$x$ &\hspace{1.0cm} $p$ &\hspace{1.0cm} $T_c$ &\hspace{1.0cm} $\Delta T_c$ \\
\noalign{\smallskip}\hline\noalign{\smallskip}
YBCO &\hspace{1.0cm}0.15 &\hspace{1.0cm} 0.006  &\hspace{1.0cm} 0 &\hspace{1.0cm} - \\
YBCO &\hspace{1.0cm}0.45 &\hspace{1.0cm} 0.081  &\hspace{1.0cm} 35 &\hspace{1.0cm} 2-3 \\
YBCO &\hspace{1.0cm}0.60 &\hspace{1.0cm} 0.115  &\hspace{1.0cm} 61 &\hspace{1.0cm} 1-3 \\
YBCO &\hspace{1.0cm}0.90 &\hspace{1.0cm} 0.165  &\hspace{1.0cm} 92 &\hspace{1.0cm} 1 \\
YBCO &\hspace{1.0cm}1.00 &\hspace{1.0cm} 0.190  &\hspace{1.0cm} 90 &\hspace{1.0cm} 1 \\
LSCO &\hspace{1.0cm}0.00 &\hspace{1.0cm} 0      &\hspace{1.0cm} 0 &\hspace{1.0cm} - \\
LSCO &\hspace{1.0cm}0.04 &\hspace{1.0cm} 0.04   &\hspace{1.0cm} 0 &\hspace{1.0cm} - \\
LSCO &\hspace{1.0cm}0.10 &\hspace{1.0cm} 0.10   &\hspace{1.0cm} 26 &\hspace{1.0cm} 1.5 \\
LSCO &\hspace{1.0cm}0.15 &\hspace{1.0cm} 0.15   &\hspace{1.0cm} 37 &\hspace{1.0cm} 1.0 \\
LSCO &\hspace{1.0cm}0.20 &\hspace{1.0cm} 0.20   &\hspace{1.0cm} 28 &\hspace{1.0cm} 3.0 \\
LSCO &\hspace{1.0cm}0.30 &\hspace{1.0cm} 0.30   &\hspace{1.0cm} 0 &\hspace{1.0cm} - \\
\noalign{\smallskip}\hline\hline
\end{tabular}
\end{table}

Despite the use of hard x-ray excitation to maximize the bulk
sensitivity of our measurements, it is still appropriate to generate a clean
surface prior to measurement, and so the single crystals were
cleaved in a preparation chamber under a vacuum of
10$^{-7}$-10$^{-9}$ mbar at room temperature by means of removal of
a top-post immediately before an in-vacuo transfer to the analysis
chamber (base pressure 3x10$^{-10}$ mbar). Only the La$_2$CuO$_4$
and La$_{1.7}$Sr$_{0.3}$CuO$_4$ samples (i.e., two out of the total
of eleven measured doping levels) proved to be uncleavable by this
method and were scraped in the preparation chamber using a diamond
file prior to the measurements. We observed that the surface
cleaning was necessary to obtain clean spectra even if hard x-rays
were used for spectroscopy.

The XPS measurements were performed at room temperature using the HIKE
experimental station at the double crystal monochromator KMC-1
beamline\cite{Schaefers2007} at BESSY, Germany. This beamline operates
on a bending magnet and the energy can be scanned in the range of
1.7 to 12 keV. The HIKE end station\cite{Gorgoi} is equipped with a
Gammadata Scienta R-4000 hemispherical electron energy analyzer
modified for high transmission and kinetic energies up to 10 keV. The
samples were studied using excitation energies of 2010 and 6030 eV
with a total energy resolution of 0.33 and 0.18 eV, respectively. The
measurements were taken at grazing incidence of the x-ray beam,
while the axis of the spectrometer lens was, within a few degrees,
normal to the sample surface. In order to vary surface sensitivity,
off-normal emission spectra were also collected in some cases. Energy
referencing was carried out with a combination of measurements of
the Fermi cutoff and Au 4$f$ levels of a scraped Au foil, held in
electrical contact with the samples. The accuracy (and
reproducibility) of the absolute and relative energy scales is of
the order of 50 meV.

\section{Results}

\subsection{YBa$_2$Cu$_3$O$_{6+x}$}

In Fig. 2, we show the Y 3$d$ core level spectra of
YBa$_2$Cu$_3$O$_{6+x}$ for the available $x$ values. The
spectra collected using 2010 eV photons in normal emission geometry
are shown in the left panel of the figure. All the spectra exhibit
a two peak structure typical of 3$d$ core levels with spin-orbit
splitting. The lines superimposed on the experimental spectra are
the result of a fit comprising two peaks shown by the thin blue
lines below each spectrum. The intensity ratio of the two component
peaks is close to 3/2 as expected from the multiplicity of the
spin-orbit split features. The component peaks in YBa$_2$Cu$_3$O$_{6.15}$
are symmetric and become asymmetric with increasing oxygen
concentration. This is the result either of more than one
electrostatic/screening environment for the Y ions upon hole doping
of the CuO$_2$ planes between which they are sandwiched, or via a
mechanism involving coupling of the core hole to a continuum of
electron-hole excitations of the charge carriers in the CuO$_2$ planes.~\cite{Doniach1970}
\begin{figure}
 \vspace{-2ex}
\includegraphics [scale=0.45]{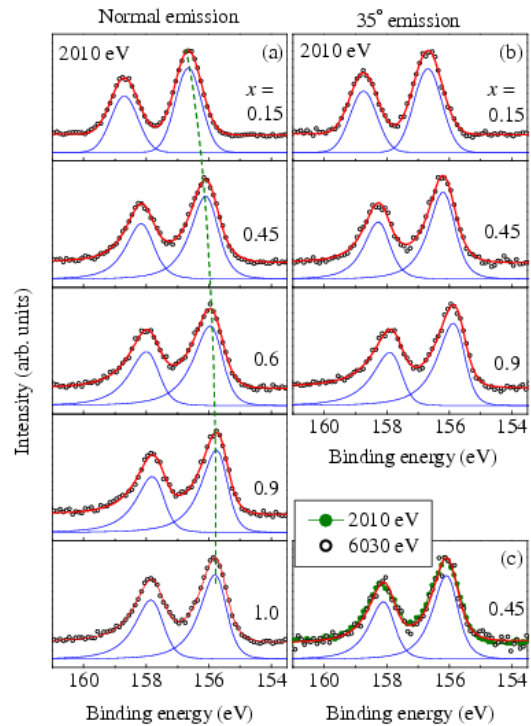}
\vspace{-8ex} \caption{(color online) Y 3$d$ core level spectra of YBa$_2$Cu$_3$O$_{6+x}$ for
various O concentrations. (a) normal emission spectra collected with
h$\nu$=2010eV. (b) analogous spectra collected
at the same photon energy but at 35$^o$ emission  with respect to
surface normal. (c) normal emission spectra collected with
2010 eV photons (solid circles) and 6030 eV photons (open circles)
for $x$=0.45. The red solid lines superimposed on the
experimental spectra are the total fit results and the thin blue
lines represent the spin-orbit split components of the fit.}
 \vspace{-2ex}
\end{figure}

In addition to this increase in asymmetry, increasing oxygen doping
also leads to a clear intensity shift of the spectral energy positions toward
lower binding energies. Between $x$=0 and 1 this shift amounts to 
$\Delta\epsilon$=0.8 eV. The analysis and discussion of this shift will
form the focus of Sec. IV B, dealing with the
chemical potential shifts as a function of doping. Readers for whom the description
of the data from the different core levels --- in terms of spectral form
and the origin of the feature --- would form an activation barrier, are free to jump directly to
Sec. IV B and consider Fig. 15 which shows the core level shift data vs. doping.

In order to test whether there is a significant surface component in
the Y 3$d$ core level spectra, we have also measured the spectra at
an emission angle of 35$^o$ with respect to the surface normal and
show these data in Fig. 2 (b). If we would take
the usual 'grazing' emission angles of 70$^o$ or more, for the high
photon energies used here, the signal would become impracticably small
(due to the mismatch between the penetration depth of the x-rays and
the inelastic mean free path length of the photoelectrons). Nevertheless, 
even the moderate emission angle of  35$^o$ used here
is sufficient to make the experiment more surface
sensitive. It is evident from Fig. 2(b) that the spectral line shapes
and energy positions of the Y 3$d$ features are essentially
unchanged compared to those found for normal emission. This would
suggest an absence of a surface-related Y $3d$ component. We have
tested this hypothesis by measuring the Y $3d$ level for $x$ = 0.45
with 6030 eV photons in the normal emission geometry, and plot the
hard x-ray data (solid circles) together with those recorded using
$h\nu$=2010 eV photons (open circles) in 
Fig. 2 (c). The spectra are identical, establishing that the Y 3$d$ core
level profile is devoid of surface-related features, and thus that
the $h\nu$=2010eV data can be considered as wholly bulk-representative.

Moving on from the Y 3$d$ to the Ba 3$d$ and 4$d$ spectra, we show
the O-doping dependent data set in Fig. 3. Ba 4$d$ spectra collected
at normal emission using $h\nu$=2010 eV are shown in Fig. 3(a).
Employment of high spectral resolution and good quality crystals
reveals distinctly separated features in contrast to earlier
data.\cite{Fowler1990,Frank1991,Teterin1993} The dominant lines at low binding energy clearly shift to
lower binding energies with increasing oxygen concentration. The
energy shift observed for the Ba core level lines is significantly
larger than that observed in the Y 3$d$ spectra shown in Fig. 2. For the Ba case the shift is
$\Delta\epsilon$=1.2 eV between $x$=0 and 1. Finally, we mention that for the shallow 
Ba $5p$ core level excitations (not shown) the same shift has been detected as for the Ba $4d$ levels. 

\begin{figure}
 \vspace{-2ex}
\includegraphics [scale=0.45]{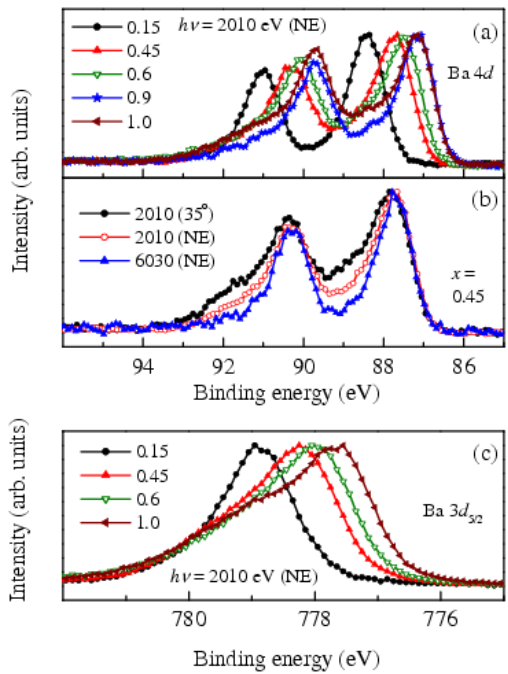}
\vspace{-8ex} \caption{(color online) (a) Ba 4$d$ core level spectra
YBa$_2$Cu$_3$O$_{6+x}$ for various $x$ values collected at normal emission with
h$\nu$=2010eV. (b) Ba 4$d$ spectra from YBa$_2$Cu$_3$O$_{6.45}$,
collected at 35$^o$ emission with 2010 eV photons, and at normal
emission with both 2010 and 6030 eV excitation energies. (c) A zoom
in on the Ba 3$d_{5/2}$ spin-orbit split component for various
values of $x$ collected at normal emission with h$\nu$=2010eV.}
 \vspace{-2ex}
\end{figure}

The Ba 4$d$ spectra exhibit distinct signatures of
multiple features in contrast to their Y 3$d$ counterparts. 
In order to investigate the origin of the clear shoulders visible on
the spin-orbit split doublets, we show the Ba 4$d$ spectra for
YBa$_2$Cu$_3$O$_{6.45}$ collected at 35$^o$ emission with 2010 eV
photon energy, and at normal emission with $h\nu$=2010 and 6030 eV
in Fig. 3(b). In this order, the bulk sensitivity of the technique
increases. It is clear that in addition to the sharp spin-orbit
split features around 87.7 and 90.5 eV binding energies, significant
intensity grows gradually around 88.7 eV and 91.5 eV binding
energies upon increasing the surface sensitivity of the measurement.
Consequently, these additional high binding energy features can be
attributed to the core level spectra from Ba ions situated at the
cleavage surface. This trend has also been observed for crystals
with other O-doping levels studied in the same manner, spanning
right across the higher doping range. The Ba 3$d_{5/2}$ spectra
- of which a portion is shown in Fig. 3(c) - also exhibit similar
behavior with respect to the probing depth of the measurement. Figure
3(c) is also well suited to illustrate that the Ba core level line
shape --- for a given measurement geometry --- also depends on the O
concentration, changing from being essentially symmetric for $x$ =
6.15 to clearly two-component for $x$ = 7.0.

\begin{figure}
 \vspace{-2ex}
\includegraphics [scale=0.45]{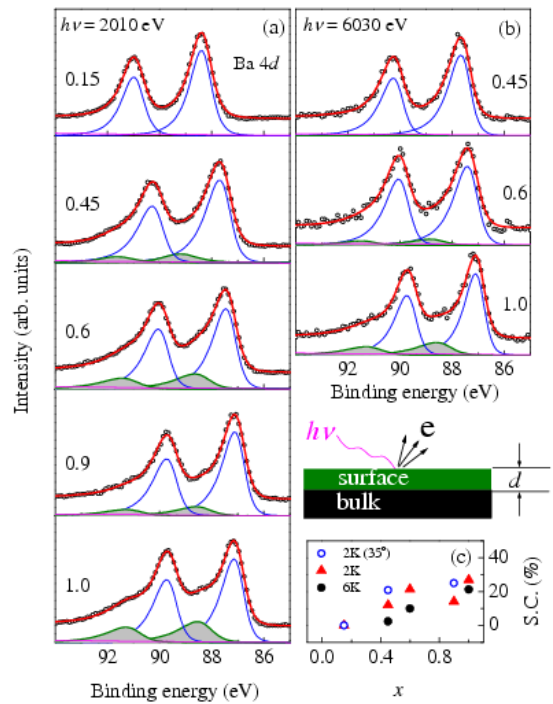}
\vspace{-8ex}
 \caption{(color online) (a) fits to the Ba 4$d$ core level spectra of 
YBa$_2$Cu$_3$O$_{6+x}$ for various $x$ values
collected at normal emission with 2010 eV photon energy. (b) Fits to
the Ba 4$d$ spectra collected at normal emission with 6030 eV
photons. (c) The surface contribution extracted from the fits, as a
function of $x$, whereby the shorthand notation 2K
and 6K has been used to designate the use of 2010 and 6030 eV
photons, respectively. S.C. represents surface contribution.}
 \vspace{-2ex}
\end{figure}

In order to gain a full understanding of the surface-bulk
differences seen in the Ba 4$d$ spectra, we have fitted the
experimental data using asymmetric line shapes, the results of which are
shown in Fig. 4. It is evident that one can simulate the Ba 4$d$
spectra of YBa$_2$Cu$_3$O$_{6.15}$ collected at 2010 eV and 6030 eV using
one pair of spin-orbit split features. This indicates that the
surface-bulk difference felt at the Ba site in this undoped
composition is negligible. Further, we note that in every
composition, the change in surface sensitivity of the technique
leads to significant change in the intensity ratio of the high to
low binding energy features, thus clearly establishing that the
spin-orbit split feature at higher binding energies is indeed from
the sample surface and the lower binding energy features are from
the bulk. 
In such a case photoemission intensity can be expressed as: $$I(\epsilon) = [1 -
e^{-d/\lambda}]I^s(\epsilon) + e^{-d/\lambda} I^b(\epsilon)$$ where
$d$, $\lambda$, $I^s(\epsilon)$, and $I^b(\epsilon)$ is the depth of the surface related layer 
as shown in Fig. 4, the effective inelastic mean free path length, the surface spectrum, and 
the bulk spectrum, respectively. The first and second terms on the right hand
side of the above equation represent the surface and bulk
contributions to the photoemission spectral function. Various
XPS studies indicate that
$d/\lambda$ is about 0.5 for electrons having a kinetic energy of about
1.5 keV ($\lambda \sim$~20\AA), which is also similar to that found
in the universal curve.\cite{Maiti2005,Maiti2006,Singh2008} Since, $\lambda
\propto \sqrt{E_{kin}}$ ($E_{kin}$ is the kinetic energy of the
photoelectrons), $d/\lambda$ for Ba 4$d$ photoelectrons is about
0.44 for the spectra corresponding to 2010 eV photons and 0.25 for
the spectra corresponding to 6030 eV photons. Thus, the surface contribution
in the Ba $4d$ spectra reduces from 36\% to 22\% with
the increase in photon energy. In Fig. 4(c) we observe that the surface contribution (S.C.) 
gradually increases with increasing $x$ and becomes 25\% for $x$=1. This suggests that
the cleaved surface contains other contributions in addition to BaO layers, however,
the probability of BaO layers
at the surface enhances with increasing $x$.

The O 1$s$ core level spectra recorded for various O concentrations
are shown in Fig. 5. All the spectra exhibit three distinct features
marked in the figure as A, B and C. With the reduction in surface
sensitivity of the technique (i.e., on going from 35$^o$ emission
with 2010 eV excitation energy, to normal emission for the same
photon energy, to using 6030 eV photons at normal emission), the
intensity of the feature C reduces significantly for all the oxygen
doping levels studied. This indicates that feature C is associated
with surface oxygens. 

\begin{figure}
 \vspace{-2ex}
\includegraphics [scale=0.45]{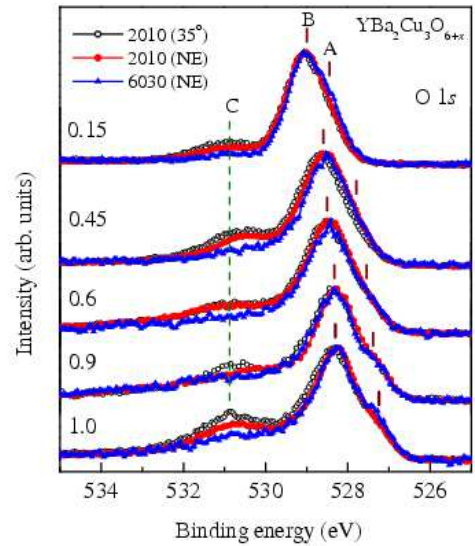}
\vspace{-20ex}
 \caption{(color online) O 1$s$ core level spectra of
YBa$_2$Cu$_3$O$_{6+x}$ for various $x$ values, collected using the
photon energies (and emission angles) indicated, whereby NE stands
for normal emission.}
 \vspace{-2ex}
\end{figure}

In order to investigate the origin of multiple features in the O
1$s$ spectra in more detail, we have simulated the experimental data
using asymmetric line shapes representing the features A, B and C.
It is evident already from a simple visual inspection of the raw
data that a minimum of three features are required to fit the whole
spectral region for all the compositions. Thus, to avoid uncertainty
in a parameterized fitting procedure, we have adopted a three peak
fit guided by a least squares optimization with respect to the raw
data. Two typical cases of fit results are shown in Fig. 6(a) and
6(b), illustrative for all the fits carried out. The doping
dependent change in the areas of the various features is shown in
Fig. 6(c). Firstly, we observe that the intensity of feature A
remains almost unchanged in all the spectra. Feature C is related to
the surface oxygens as argued above from its surface sensitivity
dependence, and gradually increases with increasing $x$. The
intensity of feature B also gradually increases with increasing
$x$, which indicates a role for the O(1) chain oxygen atoms in the
origin of this core level feature.

\begin{figure}
 \vspace{-2ex}
\includegraphics [scale=0.45]{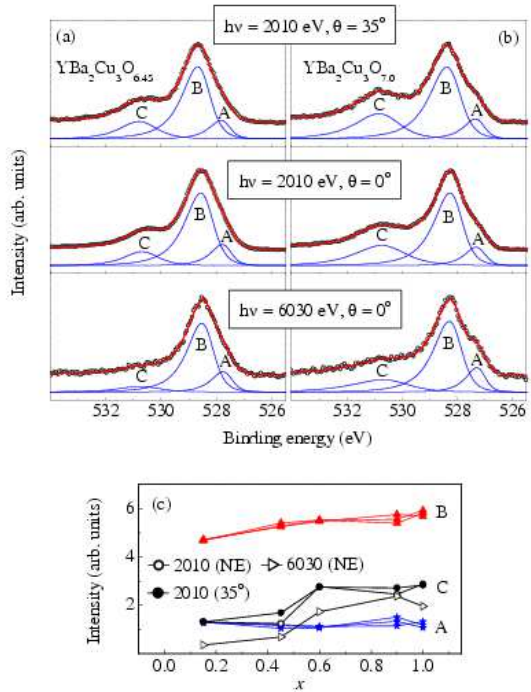}
\vspace{-8ex}
 \caption{(color online) O 1$s$ core level spectra of (a)
YBa$_2$Cu$_3$O$_{6.45}$ and (b) YBa$_2$Cu$_3$O$_{7.0}$ collected
under conditions giving differing surface sensitivity. For the
panels containing spectra: from top to bottom the bulk sensitivity
is increasing. The red (blue) lines represent the simulated total
(partial) spectral functions. (c) The areas (i.e. intensities) of
the features A, B and C as a function of $x$.}
 \vspace{-2ex}
\end{figure}

The binding energy of the various oxygen atoms in YBa$_2$Cu$_3$O$_{7.0}$ 
have been studied in several
calculations\cite{krakauer1988,Zaanen1989,Pickett1989} and using XAS.\cite{Nucker1995} 
From these earlier investigations it
emerges that O(4) (see Fig. 1) has the lowest binding energy. The
binding energies for the couple O(2) \ and O(3) and lastly O(1)
are about 0.7 eV and 0.3 eV, respectively relative to O(4).
Therefore, feature A in Figs. 5 and 6 which appears at the lowest
binding energy can be attributed to O(4) from the bulk of the
crystal. No distinct signatures of the planar oxygen ions O(2) and
O(3) and the chain oxygen O(1) could be resolved in the spectra and
thus, all three of these O sites contribute to the intensity of peak
B. This is perfectly reasonable with regard to the data shown in
Fig. 6 (c), in which the intensity of feature B grows gradually with
$x$, as can be expected if it is partially O(1)-related.

In the final part of this subsection, we present the Cu
2$p$ core level spectra. In the case of YBa$_2$Cu$_3$O$_{6+x}$, a simple attribution
of each spectral feature to a particular atomic site is severely
complicated by a number of factors. Firstly, the existence of two
types of Cu site --- plane and chain --- which take on differing
valencies and have differing dimensionalities as a function of O
content. Secondly, it is well known that the strong final state
effects visible in the Cu $2p$ spectra of (nearly) divalent cuprate
networks are both dependent on the doping level and dimension of the
Cu-O structure in question.\cite{Boske1998,Maiti1997,Maiti1998,Koitzsch2002}

\begin{figure}
 \vspace{-2ex}
\includegraphics [scale=0.45]{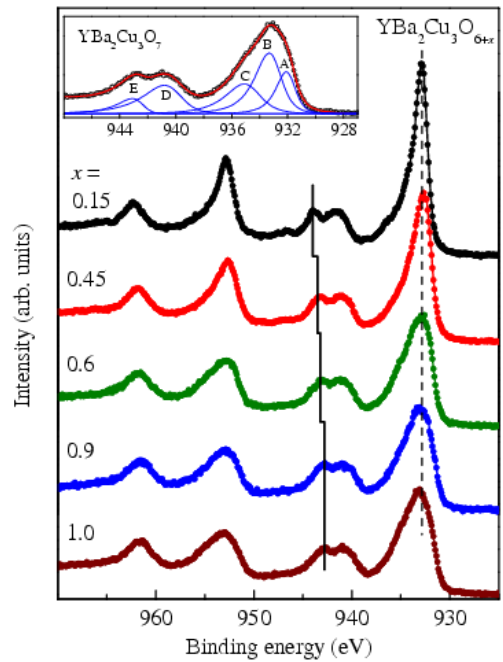}
\vspace{-8ex}
 \caption{(color online) Cu 2$p$ spectra of YBa$_2$Cu$_3$O$_{6+x}$ for
 various $x$ values collected at normal emission
using 2010 eV photon energy.}
 \vspace{-2ex}
\end{figure}

The Cu 2$p$ spectra collected using 2010 eV photons at normal
emission are shown in Fig. 7. The data exhibit four spectral
features arising from the spin-orbit split 2p$_{3/2}$ and 2$p_{1/2}$
levels and their satellites. In Fig. 7, the data have been
normalized to the area under the 2p$_{3/2}$ satellite. Since the
spectral line shape of Cu 2$p_{1/2}$ is very similar to the line shape
of Cu 2$p_{3/2}$, we concentrate on the Cu 2$p_{3/2}$ features in
the following as shown in Fig. 8. In divalent cuprates, the main
peak around 933 eV binding energy represents the well-screened final
state $\mid$\underline{2p}$3d^{10}$\underline{L}$\rangle$, whereby
the intrinsic hole in the 3$d$-shell has been pushed away to the
ligand atoms (in this case O) due to the positive core-hole
potential.\cite{Larsson1975} The satellite feature, appearing in the
binding energy range of 938-947 eV corresponds to
$\mid$\underline{2p}$3d^{9}\rangle$, and is often referred to as the
poorly-screened final state. The fact that, for the main line, the
core-hole screening takes place via charge transfer to the ligands
(and also thereby to neighboring CuO$_4$-plaquettes
\cite{vanVeenendaal1993}) means that an analysis of this kind of data
offers the opportunity not only to determine parameter values for
the model Hamiltonians for the systems in question, but also to
study the dynamics of charge transfer in cuprate
networks.\cite{Boske1998,Maiti1997,Maiti1998,Golden2001,Koitzsch2002}

\begin{figure}
 \vspace{-2ex}
\includegraphics [scale=0.45]{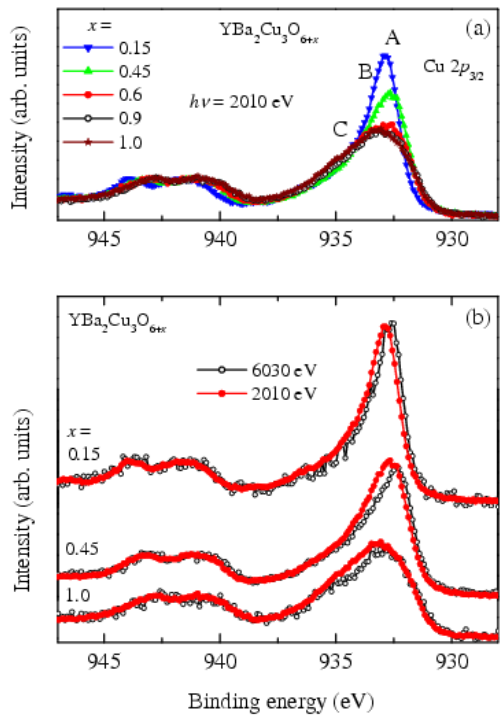}
\vspace{-8ex}
 \caption{(color online) (a) Cu 2$p_{3/2}$ spectra from
YBa$_2$Cu$_3$O$_{6+x}$ for various $x$ values recorded with 2010 eV
photons. (b) Analogous data for the doping levels $x=$ 6.0, 6.45 and
7.0 collected with h$\nu$=2010 eV and 6030 eV.}
 \vspace{-2ex}
\end{figure}

Fig. 8(a) shows an overlay of a zoomed region of the spectra showing
the Cu 2$p_{3/2}$ features for the all doping levels studied. 
Starting with YBa$_2$Cu$_3$O$_{6.15}$, as outlined above, Cu(1) is monovalent and Cu(2) is divalent.
XPS measurements on the monovalent Cu$_2$O compound~\cite{Nucker1987} show only the well-screened peak
and no satellite since the $3d$ shell is already filled in the initial state.
Thus the satellite peak for $x$=0 is only due to the Cu(2) atoms in the planes
while the well-screened peak is caused by a sharp Cu(1) peak and a broader
Cu(2) peak from the planes. The broadening of the Cu(2) peak is related to various
screening channels, more localized ones which come at higher binding energies
and more delocalized ones which appear at lower binding 
energy.~\cite{vanVeenendaal1993,Koitzsch2002,Karlsson1999}
Increasing $x$ leads to an oxidation of the Cu(1) atoms thus increasing the number of divalent Cu at the expense of
monovalent Cu. Thus the sharp feature within the well-screened peak is reduced and the satellite 
intensity increases due to 
divalent Cu in the chains. With increasing $x$, similar to the Y and Ba core levels discussed above, 
the binding energy of the satellite moves to lower binding energies. In the context of the chemical 
potential shift, this will be discussed in Sec. IV B. The shift of the well-screened peak
as a function of $x$ is essentially impossible to evaluate, since various screening channels lead to 
a complicated structure and the screening conditions also probably change for different $x$ values.

\subsection{La$_{2-x}$Sr$_x$CuO$_4$}

In Fig. 9, we show the La 3$d_{5/2}$ spectra from the different doping levels collected using 2010 eV
and 6030 eV photon energies. Each
spectra exhibits two features corresponding to the well-screened main
peak at about 833 eV and poorly screened satellite peak at about
837.5 eV binding energies.~\cite{Fuggle1983} The line shape of the features does not
change with the change in incident photon energy. In order to verify
this, we have superimposed the high photon energy spectra (solid
line in Fig. 9(a)) by shifting the spectra by about 0.2 eV toward
higher binding energies. Evidently the two spectra are very similar.
Since the bulk sensitivity of the 6030 eV spectra is significantly
higher than the 2010 eV spectra, the above observation suggests that
the bulk La 3$d$ features have lower binding energies compared to
the surface La 3$d$ spectra. This is very similar to the
observations in the Ba core level spectra in YBa$_2$Cu$_3$O$_{6+x}$ samples. This also
indicates that the sample surface contains significant amount of La.

\begin{figure}
 \vspace{-2ex}
\includegraphics [scale=0.45]{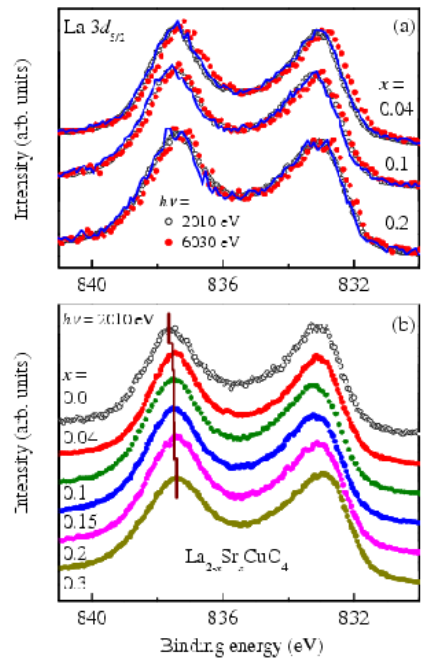}
\vspace{-2ex}
 \caption{(color online) (a) La 3$d_{5/2}$ feature and its satellite of 
La$_{2-x}$Sr$_x$CuO$_4$ for various $x$ values are
shown for 2010 eV (open circles) and 6030 eV (solid circles) photon
energies. The solid line represent the 6030 eV spectra shifted by
about 0.2 eV towards higher binding energies. (b) La 3$d_{5/2}$
spectra using 2010 eV photons for different values of $x$.}
 \vspace{-2ex}
\end{figure}

In Fig. 9(b), we compare the La 3$d_{5/2}$ features corresponding to
different doping levels. The vertical lines indicate a guide to the
energy shift. The spectra at different doping levels exhibit a
small shift in peak positions indicating finite chemical
potential shifts as observed in earlier studies. In order to verify
this, for a La core level with lower intrinsic life time broadening 
we show the La 4$d$ spectra in Fig. 10. Each spectrum exhibits
multiple features. To delineate the origin of various features, we
have fit the experimental spectra using a set of doublets, where each
doublet represents a main peak and corresponding satellite feature
as seen in the La 3$d$ spectra. All the spectra could be simulated
remarkably well by two doublets with an intensity ratio of 2:3,
which is similar to that expected for the spin-orbit split 4$d$
features. Interestingly, the peak positions estimated in this way
reveal an energy shift as a function of doping concentration
consistent with the La 3$d$ spectra.

\begin{figure}
 \vspace{-2ex}
\includegraphics [scale=0.45]{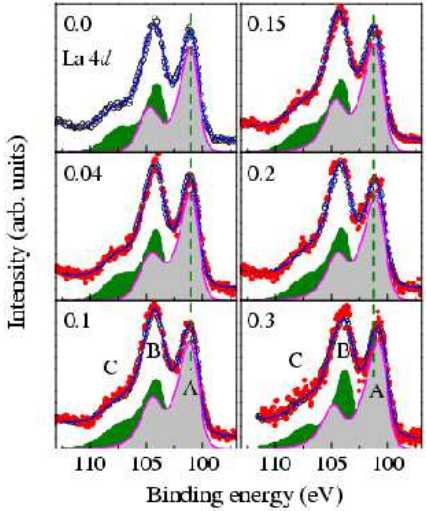}
\vspace{-16ex}
 \caption{(color online) La 4$d$ spectra of La$_{2-x}$Sr$_x$CuO$_4$ for various values of $x$.
The open and closed circles represent spectra collected using 2010
eV and 6030 eV photon energies. The solid line superimposed on the
experimental data represent the fit comprising a pair of spin-orbit split doublets, each of which
is shown by a green/gray shaded surface}
 \vspace{-2ex}
\end{figure}

The O 1$s$ spectra collected for
various compositions and different bulk sensitivities are shown in
Fig. 11. Each spectrum exhibits a two peak structure as demonstrated by fits displayed 
in Fig. 12. The most intense
feature appears at about 528.5 eV binding energy. A change in $x$
leads to a small shift in the peak positions as observed
in the La $3d$ and $4d$ signals. The feature at about 531 eV
binding energy, often attributed to an impurity feature, reduces
significantly with increasing bulk sensitivity of the
technique. However, this feature does not vanish
even if the surface is well prepared by vacuum cleaving of high
quality single crystals and hard x-ray excitation is used. Thus, this again questions its relation
to impurities, raising the possibility that it is an intrinsic feature due to surface oxygens as
also observed in YBa$_2$Cu$_3$O$_{6+x}$ samples [see Fig. 12(c)].

\begin{figure}
 \vspace{-2ex}
\includegraphics [scale=0.45]{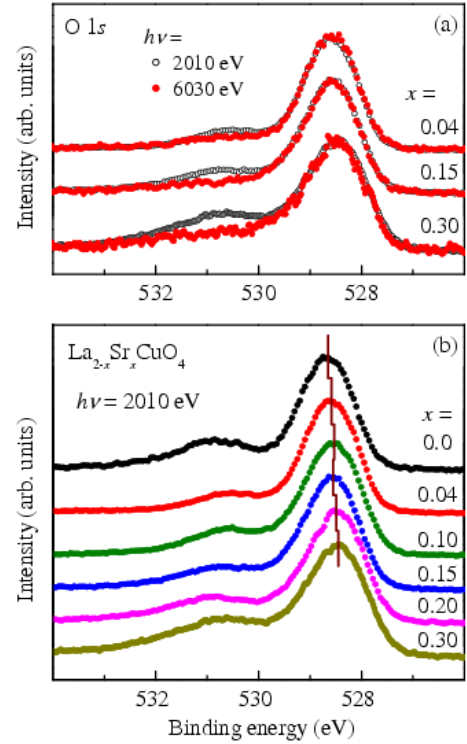}
\vspace{-2ex}
 \caption{(color online) (a) O 1$s$ core level spectra of La$_{2-x}$Sr$_x$CuO$_4$ 
for various $x$ values, corresponding to
2010 eV and 6030 eV photon energies are shown by open and solid
circles respectively. (b) 2010 eV spectra for different values of
$x$ reveal an energy shift with increasing $x$.}
 \vspace{-2ex}
\end{figure}

\begin{figure}
 \vspace{-2ex}
\includegraphics [scale=0.45]{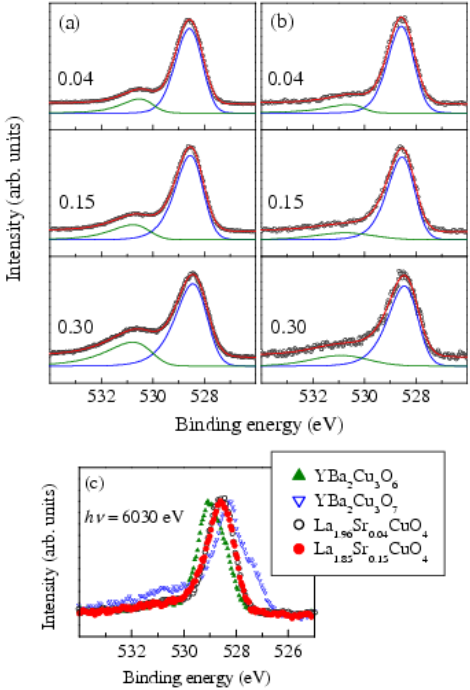}
\vspace{-8ex}
 \caption{(color online) fit to the O 1$s$ features of La$_{2-x}$Sr$_x$CuO$_4$ 
for various $x$ values obtained at (a)
2010 eV and (b) 6030 eV photon energies exhibiting two distinct
features in all the cases. (c) The peak position and profiles of the O 1$s$
spectra in La$_{2-x}$Sr$_x$CuO$_4$ are compared to those of the two 
YBa$_2$Cu$_3$O$_7$ end members.}
 \vspace{-2ex}
\end{figure}

In Fig. 13, we show the Cu 2$p_{3/2}$ signals corresponding to 2010
eV incident photon energies for different values of $x$. As described above
for the case of YBa$_2$Cu$_3$O$_{6+x}$
the satellite feature corresponds to the
non-screened excitation. Interestingly, similar to the La 3$d$ and 4$d$ and 
O 1$s$ core level spectra a small shift in binding energy of the satellite feature is detected with
the change in $x$ (see Fig. 13).

\begin{figure}
 \vspace{-2ex}
\includegraphics [scale=0.45]{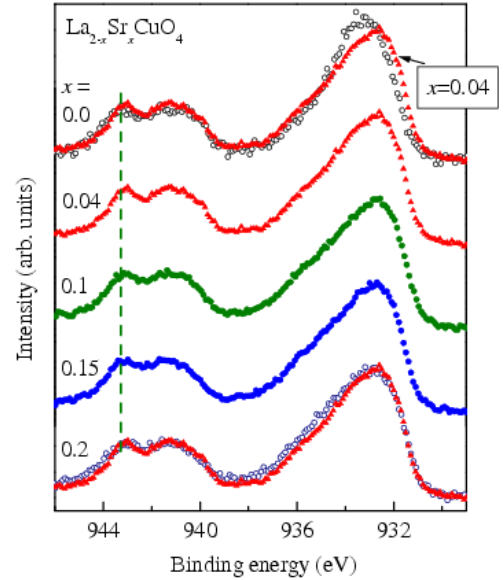}
\vspace{-12ex}
 \caption{(color online) Cu 2$p_{3/2}$ spectra of La$_{2-x}$Sr$_x$CuO$_4$ for various values
of $x$ measured with a photon energy of 2010 eV. While the satellite position shows only a small shift, the main peak exhibits
significant change. The data for x=0.0 and 0.2 are superimposed with data for x=0.04 (red triangles).}
 \vspace{-2ex}
\end{figure}

In Fig. 14 we compare the spectra obtained with
2010 eV and 6030 eV photon energies. The spectra are
normalized to the intensity of the satellite features. In this case there is no
energy shift visible in the satellite feature. In addition, the satellite features remain
unaffected even though the technique becomes significantly bulk
sensitive at $h\nu$=6030 eV, establishing again the local character of this feature.

\begin{figure}
 \vspace{-2ex}
\includegraphics [scale=0.45]{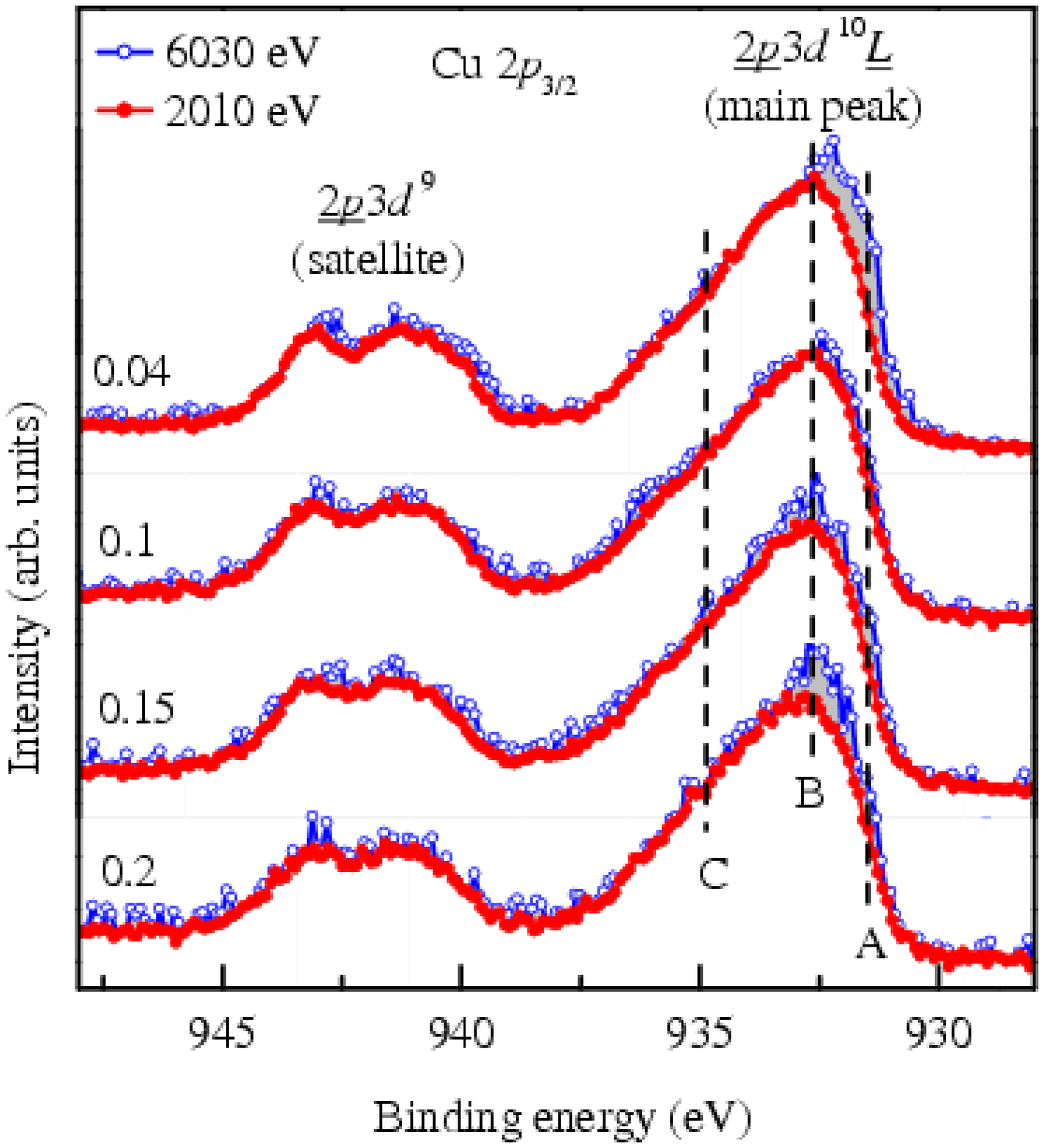}
\vspace{-12ex}
 \caption{(color online) Cu 2$p_{3/2}$ spectra of La$_{2-x}$Sr$_x$CuO$_4$ 
for various $x$ values collected using 2010
eV (solid circles) and 6030 eV (open circles) photons. A distinct
signature of three features A, B and C is visible in the spectra of the main line region.}
 \vspace{-2ex}
\end{figure}

In contrast the main peak exhibits significant redistribution of spectral weight
with increasing bulk sensitivity. Three features can be
identified in the spectra and are marked with A, B, and C in Fig. 14.
Feature C represents locally screened final states (whereby the ligand hole
resides in the CuO$_4$ plaquette containing the core hole) and remains almost unchanged in
both the spectra. Features A and B, which are caused by more delocalized screening channels
are changing when changing the surface sensitivity in the measurements, attesting to different Cu
environments in the surface and bulk regions of the sample. 

\section{Discussion}

\subsection{Surface electronic structure}
At the beginning of this discussion on the electronic structure of cleaved
cuprate crystals (which starts by dealing with the case of YBa$_2$Cu$_3$O$_{6+x}$), 
we emphasize that in the present XPS experiments we have no spatial
resolution. The diameter of the photon beam is of the order of 500 $\mu$m. Thus 
we obtain only averaged information from the surface which
may, on a microscopic scale, possess differing terminations.
On the other hand, due to the possibility of using photon energies between 2010 and 6030 eV
and by using different emission angles, depth sensitive information has been obtained, which can be used
to differentiate between the surface and bulk electronic structure. 
In previous work on cleaved YBa$_2$Cu$_3$O$_{6+x}$ single crystals (see e.g. Ref.~\onlinecite{Brundle1993}) it was
pointed out that core level excitations (except Cu ${2p}$) have a low binding energy component which represents
bulk properties and high binding components which are caused by surface contaminations
and by intrinsic surface contributions. Unfortunately the two high binding energy components appear almost at the
same energy.

As pointed out in Sec. III A, we see core level excitations from atoms where spectra 
show no sensitivity to the changes in the probing depth of the experiment, and
others which show a dependence. Examples for the latter are the Ba and O core level
excitations, while the former type is illustrated by the core lines from Y and the poorly 
screened final state satellite feature for Cu.

The lack of a high binding energy peak in the Y $3d$ spectra enables us to conclude that 
the portion of the surface terminations containing Y layers is negligible.
Furthermore, we have never observed any spin-orbit doublet for Y shifted by 1.5 eV to higher
binding energies as has been detected in air-exposed samples.~\cite{Brundle1993} 
This indicates that we do not have any
Y-related surface contamination.
The slight asymmetry of the lines at higher doping levels can be fully understood by
a Doniach-Sunjic like screening of the core holes by the metallic CuO$_2$ layers which are 
in close proximity to the Y site.

For the Ba $3d$ and $4d$ core level excitations we observe only one spin orbit
doublet at low O concentrations, with symmetric lines which we ascribe to Ba atoms in the bulk. At higher
O concentrations we see shoulders at higher binding energies. The question is whether this
second component is related to an intrinsic surface component or to surface contamination.
In a previous study, air exposed samples show a strong high energy component shifted by 1.5 eV 
compared to the bulk component.~\cite{Brundle1993}
This could indicate that the high-energy component in the spectra may be caused by carbonate
or hydroxide formation due to reaction with species in the residual vacuum. On the other hand,
we see no increase of the high binding energy component as a function of time. Therefore it is
unlikely that this component is related to contamination due to reactions with
a background pressure in the 10$^{-10}$ mbar range, or during the brief interval between cleavage at
ca. 10$^{-8}$ mbar and transfer to the analysis chamber. 
We thus assign the high binding energy component
to an intrinsic surface layer and thus conclude that Ba must be in the terminating layer.
From our measurements we cannot obtain detailed information on the possibly reconstructed
BaO-CuO-BaO complex at the crystal termination after cleavage. 
At present it is also less clear why the surface layer in the crystals 
with low O concentration does
not produce the high binding energy component, although a connection to the
population of the O(1) sites in the chains is a clear possibility. From the present measurements it follows 
that the cleavage plane or the reconstruction that may follow cleavage
probably changes between low $x$ values and high $x$ values. The spectra
also indicate that for low $x$ values there are no charged surface layers (which would
produce a high binding energy component 
in the XPS spectra) while for higher doping levels, a charged surface layer exists.
The latter results are in line with the detection of overdoped surface layers in
ARPES spectra of YBa$_2$Cu$_3$O$_{6+x}$ for higher $x$ values.~\cite{Nakayama2007,Zabolotnyy2007,Hossain2008} 

The O $1s$ spectra are also composed of a high binding energy component 
and a low binding energy component with a clear fine structure in the form of a shoulder at low 
binding energy.
In Sec. III A, the signal below 530 eV was ascribed to bulk O ions, with O(4) giving
rise to the low binding energy shoulder and the other three sites adding up to 
give the strongest feature at ca. 529 eV. 
At high photon energy ($h\nu$=6030 eV) and in particular for small $x$ values
the intensity of the high binding energy component (binding energy $>$530 eV) is rather small. This indicates 
that using similar arguments as those discussed in the case of the Ba core levels,
this part of the O $1s$ spectrum is related to an intrinsic surface contribution.
 
In the Cu $2p$ spectra, upon changing the surface sensitivity, no changes in the satellite
features could be detected. This is in line with the interpretation of this feature in terms
of atomic like excitations. Small differences in the main lines are observed which could reflect
small differences in the screening channels between the bulk and the surface. On the other hand, we
point out that the changes of the main line  of YBa$_2$Cu$_3$O$_{6+x}$ upon changing the photon energy  are 
less pronounced than in La$_{2-x}$Sr$_x$CuO$_4$
and, as will be discussed in a moment, is even less pronounced than in Nd$_{2+x}$Ce$_x$CuO$_4$. 

In the following we discuss the spectra of La$_{2-x}$Sr$_x$CuO$_4$. The common view is that
cleavage occurs between the LaO layers, thus forming LaO terminated surfaces. 
Interestingly, in the La core level spectra
taken with different surface sensitivities, no surface component could 
be detected. Only a small shift of about 200 meV
could be detected between spectra taken with $h\nu$ = 2010 eV and 6030 eV. 
This shift may be caused by a Madelung potential
at the La sites that may be different for the surface layer and the bulk.

In principle, the bulk component in the O $1s$ spectra at low binding energy 
should be a superposition of two lines stemming from
the CuO$_2$ plane O atoms and the apical O sites. Although we are performing high resolution 
XPS experiments, the two sites could not be
resolved, indicating that the binding energy of the O $1s$ levels of the two sites 
is almost the same. The high energy component
in the O $1s$ spectra in La$_{2-x}$Sr$_x$CuO$_4$ shows a strong surface sensitivity, almost 
disappearing in the spectra taken with the photon energy
$h\nu$=6030 eV. This means that similar to the case of YBa$_2$Cu$_3$O$_{6+x}$, 
the high energy component is surface related. 
In the data from the cleaved crystals ($x$=0.04 to
0.2) this component is small, while for the scraped crystals ($x$=0 and 0.3) the 
high energy component is considerably larger. The enhanced roughness of the scraped
surface leads to larger surface contributions than for the cleaved surface. Thus, the larger high 
binding energy feature in the O$1s$ spectra from the scraped sample surfaces re-establishes its connection to the
surface electronic structure. 

As mentioned previously the satellite feature in the Cu $2p$ spectra do not 
change when changing the surface sensitivity of the
XPS measurements using different photon energies in keeping with its atomic-like multiplet character.
On the other hand, in the main line the low energy features A and B in Fig. 14 
increase when measured with increased bulk sensitivity. These
features are caused from more delocalized screening channels, in which the hole on the (oxygen) 
ligands moves further away, forming
Zhang-Rice singlets. In a first approximation these screening channels would come from the in-plane electronic
structure of the CuO$_2$ layers and therefore it is difficult to understand why there should be a difference
between a surface CuO$_2$ layer and a bulk CuO$_2$ layer. On the other hand such an enhancement with 
increasing bulk sensitivity was also
observed in earlier studies, in particular for Nd$_{2+x}$Ce$_x$CuO$_4$ (Ref.~\onlinecite{Taguchi2005}). 
This could indicate that an explanation of the well-screened feature solely 
in terms of in-plane screening channels is probably not sufficient. Possibly 
the total three dimensional electronic structure
has to be used in the framework of an Anderson impurity model to calculate  
the screening channels for the Cu $2p$ core hole 
to obtain a satisfactory description of 
the spectra.~\cite{Karlsson1999,Karlsson2000,Koitzsch2002}
 
\subsection{Chemical potential shift vs. doping level}

Now that we have successfully identified lines from core level excitations in 
the XPS spectra which are definitely related
to the bulk properties and which are not influenced  by surface effects, 
we can evaluate the bulk core level binding energies as a function of doping.
Since the core level binding energies are measured relative to the Fermi level, we can
extract the chemical potential as a function of hole concentration as was described
in the Introduction. In the following, we discuss our results
on the chemical potential shifts and compare them with theoretical calculations and previous experimental results
published in the literature.
  
Without further ado, in Fig. 15 we show a compilation of all of the
relevant data for our single-crystal-only, bulk sensitive study of
YBa$_2$Cu$_3$O$_{6+x}$ and La$_{2-x}$Sr$_x$CuO$_4$. The top panel shows
the shift in the binding energy of the core levels indicated
relative to that of $x$ = 0 vs. $x$ in YBa$_2$Cu$_3$O$_{6+x}$. The
bottom panel shows the binding energy shifts vs. the hole
concentration $p$ per Cu ion in the CuO$_2$ planes.
In Fig. 15(b) we also compare these data with core level shifts derived in
La$_{2-x}$Sr$_x$CuO$_4$ for various $x=p$ values. For both systems the binding energy shift values
are given relative to $p=0$. 

\begin{figure}
 \vspace{-2ex}
\includegraphics [scale=0.45]{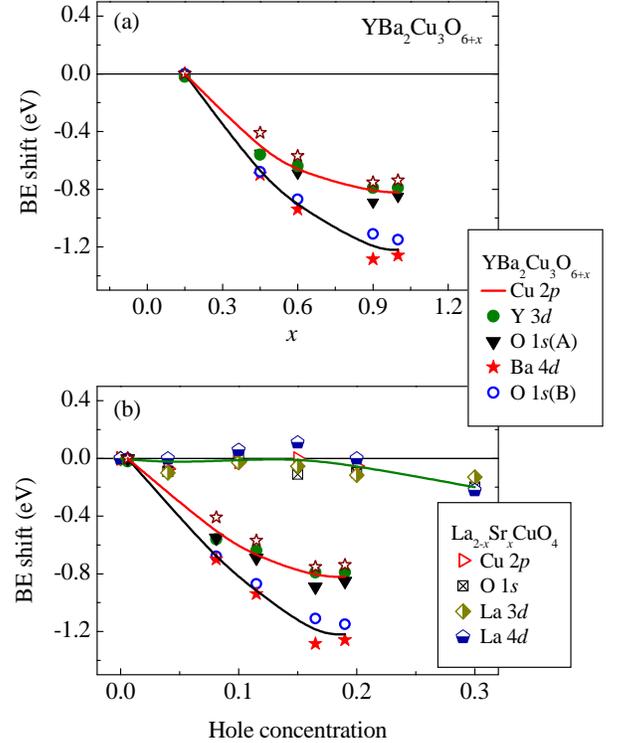}
\vspace{-8ex}
 \caption{(color online) binding energy shift of the various bulk-related core levels
lines as a function (a) of $x$ and (b) of
hole concentration in the CuO$_2$ layers 
in YBa$_2$Cu$_3$O$_{6+x}$ samples.
The analogous energy shifts for La$_{2-x}$Sr$_x$CuO$_4$ samples are also shown in (b). 
The solid lines represent a guide to the eye.}
 \vspace{-2ex}
\end{figure}
In La$_{2-x}$Sr$_x$CuO$_4$ the core level excitations from O, La, and Cu atoms show 
$\it{no}$ shift up to and including $p$=0.15. For higher $p$ values a negative shift is detected (i.e., the binding
energy of the core levels, referred to the Fermi energy is decreasing) which reaches 
a value of 0.2 eV for $p$=0.3.
Our results are in perfect agreement with the data published in Ref.~\onlinecite{Ino1997}.
The fact that the shifts $\Delta\epsilon$ are zero for $p\leq$0.15 may be interpreted by an accidental
cancellation of the four terms in Eq. (2) in this Sr concentration range. However, we
believe that such a cancellation for three different atoms in such a large concentration range
is rather unlikely. Thus we assume 
that in Eq. (2) the Madelung term, the on-site valency term, 
and the relaxation term are negligible for the particular core level features under consideration, and that
the observed shift is determined by the chemical potential shift $\Delta\mu$.
We note here that, contrary to previous investigations, we have determined the energy shift for the
Cu $2p$ lines not from the main line but from the satellite and thus avoid
'contamination' of the core level shift values with chemical shift terms
related to the changing copper valence.
For the O atoms the situation is less clear since the holes which are formed upon doping with Sr 
are situated mainly on the O sites, as a result of the formation of Zhang-Rice singlets. 
Thus, the $2p$ count on the O 
sites will be changed, and a chemical shift of the O $1s$ core level should be expected.
Apparently, the constant K in Eq. (2) is too small to enable the detection of this shift.
The calculated Madelung terms in a point charge model are of the order of 2 eV for
doping concentrations up to $x$=0.2 (see Ref.~\onlinecite{Fujimori2002}).
Opposite shifts are derived from these calculations for Cu vs. La and O core
level excitations.
On the other hand, these calculated values are probably strongly reduced by a high background dielectric
constant. This assumption is supported by the experimental fact for the core levels of the three types of atoms,
the same shift (or no shift) is observed, independent of the distance to the dopant atoms (here Sr). 
Turning to the last term in Eq. (2), it is plausible that the relaxation energies do not significantly depend
on the doping concentration because the observed core level
shifts for atoms in metallic CuO$_2$ layers and insulating LaO layers are shown to be the same.
We also emphasize that the three different types  of atoms are bonded quite differently.
For La  and O in the LaO planes a predominantly ionic bonding is a reasonable starting point,
whereas Cu and O in the CuO$_2$ planes are expected to show a strong covalent bonding component.
Thus the results for La$_{2-x}$Sr$_x$CuO$_4$ would indicate that the chemical potential
is pinned by impurity levels within the charge-transfer gap when going from the undoped insulating
system to the doped metallic system. More precisely, the inverse charge susceptibility 
$\delta\mu/\delta p$ is zero for $0\leq p\leq 0.15$.

For YBa$_2$Cu$_3$O$_{6+x}$ the results for the chemical potential shift are 
completely different to those of La$_{2-x}$Sr$_x$CuO$_4$ [see Fig. 15(b)]. 
Interpreting the measured core level shifts in terms of a 
chemical potential shift, we obtain from the data for $p<0.1$, presented in Fig. 15(b), 
an inverse charge susceptibility $\frac{\delta\mu}{\delta p}|_{p\to0}$ between 5.5 and 9.5 eV/hole
depending which core level lines one considers. 
This would support the simple semiconductor scenario in
which upon doping the Fermi level moves from the center of the gap rapidly into the valence band
(effective lower Hubbard band in a single band model). At higher hole concentration ($p\geq 0.15$) 
the shift seems to saturate.
For one group of core level excitations the maximum shift saturates at -0.8 eV while for others
a maximum shift of -1.2 eV is detected. The lower value would correspond reasonably to half of the
gap between the valence band and the upper Hubbard band in the insulating system. 
The higher saturation value is slightly
too big when compared to half of the gap. This could indicate that the values
from the Ba $4d$ and the O $1s$(B) excitations are in some way less directly related to the
(bulk) chemical potential shift. It is remarkable that these two excitations both stem from
the BaO(4) plane while the other excitations predominantly stem from the CuO$_2$-Y-CuO$_2$ complex
[except the small contribution from the O(1) atoms in the CuO(1) chains].
Having said that, it is still unclear as to why the behavior of the core level shifts should group in this manner.
A priori we would have expected that
the shifts from the ionic atoms such as Ba and Y and from the Cu $2p$ satellite would be the same.
The large shifts of the Ba core level excitations as a function of doping were also discussed in 
Ref.~\onlinecite{Brundle1993}, however, without a possible chemical potential shift being taken into account.
Both the Ba and the O(4) atoms are close to the O(1) dopant atoms
and possibly experience an additional negative shift due to a Madelung term caused by the increasing number
of O(1) atoms. Since the Cu(1) atoms contribute only 1/3 to the Cu $2p$ spectra this effect is possibly
not detected in the Cu $2p_{3/2}$ spectra. In this context one should also mention that in these
Cu $2p_{3/2}$ spectra, the Cu(1) atoms only contribute to the satellite for $p$ larger than $\approx0.3$
since for lower $p$ values they are monovalent and therefore do not form a satellite feature.

In view of the discussions above, we tentatively assign the chemical potential shift in this compound to values
measured via the Cu $2p$, Y $3d$, and O $1s$ (line A) core level excitations. We then come to
a value for the inverse charge susceptibility  $\frac{\delta\mu}{\delta p}|_{p\to0}\approx5.5$ eV/hole.

This non-zero value is in agreement with values derived in previous studies. In Ref.~\onlinecite{vanVeenendaal1993}
an inverse charge susceptibility $\frac{\delta\mu}{\delta p}|_{p\to0}\approx7$ eV/hole  
has been derived for the system
Bi$_2$Sr$_2$Ca$_{1-x}$Y$_x$Cu$_2$O$_{8+\delta}$ which saturates for $p>0.1$. The total shift between
$p=0$ and $p=0.25$ amounts to $\Delta\mu=0.8$ eV. On the other hand in Ref.~\onlinecite{Hashimoto2008} 
a `universal` inverse charge 
susceptibility  $\frac{\delta\mu}{\delta p}|_{p\to0}\approx1.8$ eV/hole has been
observed for various non-La$_{2-x}$Sr$_x$CuO$_4$ cuprates [Bi$_2$Sr$_2$Ca$_{1-x}$R$_x$Cu$_2$O$_{8+\delta}$ 
(R = Pr, Er), Ca$_{2-x}$Na$_x$CuO$_2$Cl$_2$, and 
Bi$_2$Sr$_{2-x}$La$_x$CuO$_{6+\delta}$].
Thus we are left with two puzzles (i) why is La$_{2-x}$Sr$_x$CuO$_4$ the only  p-type doped cuprate which shows
a negligible inverse charge susceptibility for small $p$ and (ii) why are the inverse 
charge susceptibilities of the non-La$_{2-x}$Sr$_x$CuO$_4$ systems so different?

To begin with we deal with the second puzzle. For the various cuprates, the energy bands near the Fermi level
originating from the CuO$_2$ planes are fairly similar.~\cite{Pickett1989} In this way we could
understand why our values $\frac{\delta\mu}{\delta p}|_{p\to0}$ and $\Delta\mu$ between $p$=0 and $p=0.2$ for
YBa$_2$Cu$_3$O$_{6+x}$ agree reasonably with the values for Bi$_2$Sr$_2$Ca$_{1-x}$Y$_x$Cu$_2$O$_{8+\delta}$ 
presented in Ref.~\onlinecite{vanVeenendaal1993}.
On the other hand, it is difficult to understand why
YBa$_2$Cu$_3$O$_{6+x}$ has such a different inverse charge susceptibility when compared with the `universal` curve
derived in Ref.~\onlinecite{Hashimoto2008}. One way to explain this discrepancy would be taking into account 
the formation of CuO$_3$ chains, only existing in YBa$_2$Cu$_3$O$_{6+x}$. On the other hand,
for small $x$ values we start with the CuO(4)$_2$ dumbbells, which upon doping, i.e., 
insertion of O(1) atoms, form more and
more fragmented chains which cause localized states, also close to the Fermi level.~\cite{Pickett1989}
Upon completely filling the chains, a Cu(1) $3d_{y^2-z^2}$-O(1) $2p_y$-O(4) $2p_z$ band is formed which is similar
to the CuO$_2$ layer bands. In addition, close to the Fermi level another chain band occurs which is caused by
a $\pi$ hybridization of O $2p$ orbitals of the O(1) and O(4) atoms. 
Both this band and the localized states formed by the
fragmented chains would be more likely to cause pinning of the chemical potential 
leading to a smaller chemical potential shift but not an increase
of the inverse charge susceptibility. Therefore it is difficult to understand why the inverse charge susceptibility
detected in YBa$_2$Cu$_3$O$_{6+x}$ in the present experiments is so much larger
than in the other cuprates studied in Ref.~\onlinecite{Hashimoto2008}. In this context we also mention that this 
large chemical potential shift in YBa$_2$Cu$_3$O$_{6+x}$ is not due to formation of metallic CuO$_3$ chains 
before the insulator metal transition
appears in the CuO$_2$ planes. XAS measurements on untwinned single crystals clearly indicate that the number of holes
in both CuO units shows almost the same doping dependence.~\cite{Nucker1995} This indicates, when considering the 
dimensionality of the two
units, that the insulator-metal transition comes first in the planes and later 
in the chains, meaning that the observed
chemical potential shift for low doping in the CuO$_2$ planes is not determined by the chains but by the planes.
Finally, in this context we mention measurements of the plasmon 
dispersion by EELS.~\cite{Nucker1989,Nakai1990,Romberg1990}
The dispersion coefficient extracted from those measurements 
yield information on the momentum dependent compressibility
of the electron liquid. For the two compounds YBa$_2$Cu$_3$O$_{7}$ and Bi$_2$Sr$_2$CaCu$_2$O$_{8}$ almost the
same dispersion coefficient has been detected which 
fits the  YBa$_2$Cu$_3$O$_{6+x}$/Bi$_2$Sr$_2$Ca$_{1-x}$Y$_x$Cu$_2$O$_{8+\delta}$
agreement  for the chemical potential shift we observe, but raises questions why Er and Pr-doped 
Bi$_2$Sr$_2$CaCu$_2$O$_{8+\delta}$ and La-doped Bi$_2$Sr$_{2}$CuO$_{6+\delta}$ appears to be different
in their carefully conducted XPS experiments.

We have confirmed in this work earlier results for a
zero inverse charge susceptibility for La$_{2-x}$Sr$_x$CuO$_4$ for small $x$ values (see Ref.~\onlinecite{Ino1997}).
In addition, in our present work we have discovered a further system which shows 
a large inverse charge susceptibility, but
which is even larger than in various other non-La$_{2-x}$Sr$_x$CuO$_4$ compounds.
As pointed out in the Introduction various models for the existence or non-existence of a finite inverse
charge susceptibility have been proposed.

Several theoretical calculations~\cite{Fisher1995,Kajueter1996,Georges1996} have predicted for a p-type
doped Mott-Hubbard insulator the appearance of a quasiparticle like resonance
at the top of the lower Hubbard band. This would give a jump of the chemical potential shift $\Delta\mu$=$U$/2. 
In a charge-transfer insulator (which can be transformed into an effective Mott-Hubbard insulator)
the shift would be half of the charge-transfer gap. These calculations also predict that the resonances move into the
gap depending on the size of $U$ relative to a critical value $U_c$. Such a shift could reduce the inverse charge 
susceptibility as experimentally observed in the
system La$_{2-x}$Sr$_x$CuO$_4$. On the other hand, in view of the similarities 
of the crystal structure and the electronic structure
of the CuO$_2$ planes in the various cuprate high-$T_c$ superconductors, 
it is very difficult to understand why the parameters
determining the energy position of the quasiparticle resonance should be so different for the various compounds.

Another reason for a vanishing chemical potential shift upon doping a Mott-Hubbard 
insulator could be a phase separation 
into a metallic and an insulating phase. Microscopic phase separation has been 
claimed to be detected in the stripe phases
in which antiferromagnetic insulating antiphase domains are separated by periodically 
spaced domain walls to which the holes
segregate (see e.g. Ref. \onlinecite{Vojta2009}). Actually, as mentioned in the Introduction,
a slave boson mean-field analysis of the three-band model resulted in a constant chemical potential up to $p$=1/8.
At this doping concentration --- corresponding to four lattice constants in
the CuO$_2$ plane --- the cuprates are most susceptible to stripe formation. Below this doping concentration,
the distance between the rivers of charge increase more and more, leaving the hole concentration
within the microscopic metallic regions (the stripes) unchanged. This could lead to a vanishing
chemical potential shift, an argument which was used for the explanation of the zero
inverse charge susceptibility observed in La$_{2-x}$Sr$_x$CuO$_4$ [Ref.~\onlinecite{Fujimori2002}]. 
Above $p$=1/8 the common view is that in
the stripe phase, the doping concentration within the stripes increases.
In this scenario based on microscopic phase separation not only the vanishing 
chemical potential shift for La$_{2-x}$Sr$_x$CuO$_4$ for small x was explained but also the reduced
$\delta\mu/\delta p$ for R$_{1-x}$A$_x$MnO$_3$ (R: rare earth; A: alkaline earth) around the 
charge ordered composition range.~\cite{Ebata2008} In addition, the fixed
chemical potential in La$_{2-x}$Sr$_x$NiO$_4$ for $x\leq$0.4 [Ref.~\onlinecite{Satake2000,Fujimori2002}] 
was also discussed in this context.

On the other hand, we see serious problems to explain the vanishing chemical potential shift in
La$_{2-x}$Sr$_x$CuO$_4$ for small $x$ on the basis of a microscopic phase separation, as, in this
compound only fluctuating stripes have been observed. Only in the systems  La$_{2-x}$Sr$_x$CuO$_4$ in which
part of La is replaced by Nd or Eu, or Sr is replaced by Ba, have static stripes been detected.
Furthermore, for $x$=0.15 we are definitely in a range with strongly interacting stripes which is  a region far from
that discussed in terms of microscopic phase separation. 
Finally, in this scenario it is very difficult to explain the difference between
La$_{2-x}$Sr$_x$CuO$_4$ and YBa$_2$Cu$_3$O$_{6+x}$ as regards the core level shifts since 
in the former system only fluctuating
stripes exist while in the latter system a nematic stripe liquid has 
even been detected experimentally for $x$=0.45 (Ref. \onlinecite{Hinkov2008}).

Another explanation of the anomalous inverse charge susceptibility in La$_{2-x}$Sr$_x$CuO$_4$ is based on the role
of the next-nearest neighbor hopping integral $t'$ in the band dispersion of cuprates. Theoretical work going
in this direction was published in Ref.~\onlinecite{Tohyama2003} predicting an increasing chemical potential
shift with increasing $t'$. This scenario is related to the van Hove
singularity close to the Fermi surface of the two dimensional electronic structure of 
the CuO$_2$ planes. This singularity
causes a high density of states close to the Fermi level which in turn may reduce the chemical potential shift.
For a small $t'/t$ ($t$ is the nearest neighbor hopping integral) causing a square 
Fermi surface, the chemical potential 
lies in the singularity and doping causes only a small shift.
Larger absolute $t'/t$ values shift the singularity below the Fermi level thus 
increasing the chemical potential shifts
caused by small doping concentrations. On the other hand, a compilation of $t'/t$ values based on band structure 
calculations~\cite{Pavarini2001}  shows that $t'/t$ for La$_{2}$CuO$_4$ is considerably larger than that of 
Ca$_{2}$CuO$_2$Cl$_2$. This would give a smaller chemical potential shift for the latter when compared with the former
in clear contradiction with the experimental findings. Moreover, the band structure calculations and also 
ARPES data indicate that in all the
cuprates $t'/t$ is finite, shifting the van Hove singularity well below the Fermi level. 
Thus a doping concentration near 0.2 is needed 
to shift the chemical potential into the sigularity which could reduce the chemical potential shift.
In a recent paper,~\cite{Hashimoto2008} therefore, a doping dependent $t'/t$ has been introduced to explain
the small $\frac{\delta\mu}{\delta p}|_{p\to0}$ in La$_{2-x}$Sr$_x$CuO$_4$.

For completeness we also mention discussions of the influence of a pseudogap on the chemical potential shift in 
doped cuprates.~\cite{Tjernberg1997,Fujimori2002} Since the observed pseudogaps in the underdoped cuprates are of the
order of 60 meV they have a negligible influence on the chemical potential shifts detected in this work. Furthermore
the existence of a pseudogap in underdoped cuprates is a generic phenomenon and cannot 
therefore explain the differences in
the chemical potential shifts between different cuprates, which were discussed above.

\section{Conclusions}

We have studied the core level spectra of La$_{2-x}$Sr$_x$CuO$_4$
and YBa$_2$Cu$_3$O$_{6+x}$ samples for different values of $x$ using
high resolution, hard $x$-ray photoemission spectroscopy with variable photon energies up
to 6030 eV. In those spectra we could separate core level features derived only from the bulk
and features which come from the surface. The latter seem
predominantly caused by intrinsic surface contributions, i.e., contributions which
are related to the cleaving of the surface which leads to reconstructions
and charge redistributions at the surface, and only a small part seems due to surface contamination. The
experimental results establish that the cleavage in
YBa$_2$Cu$_3$O$_{6+x}$ crystals for $x\geq$0.45 occurs predominantly in the BaO(1)Cu(1)O(4)$_2$ layers which 
result in charged terminations. 
For small $x$ values the cleavage planes are found to be again in the Cu(1)O(4)$_2$ layers, 
thus leading in this case to
a negligible contribution from charged surface terminations. 

Evaluating the binding energies of the core levels from the unambiguously bulk-related features
we could determine the chemical potential shifts as a function of the doping concentration in the CuO$_2$ planes.
For La$_{2-x}$Sr$_x$CuO$_4$ we could confirm previous results in the literature
indicating no shift of the chemical potential up to and including a doping concentration $x$=0.15. 
In view of a simple picture of a doped Mott-Hubbard insulator, a large jump
of the chemical potential of the order of half the gap energies, i.e., of the order of
0.8 eV is expected.
In YBa$_2$Cu$_3$O$_{6+x}$, on the other hand this expected large shift is detected
when going from the undoped material to the optimally doped high-$T_c$ superconductor.
The observed concentration dependence of the chemical potential shift is in good
agreement with that observed in Bi$_2$Sr$_2$Ca$_{1-x}$Y$_x$Cu$_2$O$_{8+\delta}$.~\cite{vanVeenendaal1993}
However, this shift is considerably larger than that in three other doped cuprate
systems presented in the literature.~\cite{Hashimoto2008} In particular the changes of the chemical potential shift
per hole, i.e., the inverse charge susceptibility for small $x$ 
values differs considerably between different (non- La$_{2-x}$Sr$_x$CuO$_4$) p-type doped cuprates.
These contradicting results are discussed in terms of different scenarios which could influence the
chemical potential shift as a function of doping: the influence of microscopic charge separation (stripe
formation), particular band structure phenomena such as the influence of the next-nearest neighbor 
hopping integral $t'$, or the 
pseudo gap. None of these scenarios yield a satisfactory explanation covering the detected experimental
results. Therefore, although we embarked on this systematic, bulk sensitive
study of the doping dependence of the chemical potential $\mu$ in
YBa$_2$Cu$_3$O$_{6+x}$ and La$_{2-x}$Sr$_x$CuO$_4$ in order to achieve a generically valid 
picture for this fundamental
property of these electron systems, we are obliged to close this paper with the call for further experimental
and theoretical work that still seems to be necessary to understand the ongoing chemical potential
puzzle in the cuprates. 
\\

\hspace{1cm}{\bf ACKNOWLEDGEMENT}\\

J.F. is grateful for discussions with M. Vojta. We acknowledge technical assistance by F. Sch\"{a}fers. 
K.M. thanks the Alexander von Humboldt Stiftung, Germany and
the Leibniz-Institute for Solid State and Materials Research, Dresden,
Germany for financial assistance.  J.F., V.H. and A.E. appreciate financial support by the 
DFG (Forschergruppe FOR 538). This work is part 
of the research program of the 'Stichting voor Fundamenteel Onderzoek 
der Materie (FOM)', which is financially supported by the 'Nederlandse Organisatie voor Wetenschappelijk 
Onderzoek (NWO)'.
This work was supported by the European Community - Research 
Infrastructure Action under the FP6 ''Structuring the European Research 
Area'' Programme (through the Integrated Infrastructure Initiative 
''Integrating Activity on Synchroton and Free Electron Laser Science'' - 
Contract R II 3-CT-2004-506008)

\bibliographystyle{apsrev}
\bibliography{YBCO_submit}

\end{document}